\documentstyle[floats,prd,aps,epsf]{revtex} 
\draft 

\begin{document}

\twocolumn[\hsize\textwidth\columnwidth\hsize\csname
@twocolumnfalse\endcsname

\def\boxit#1{\vbox{\noindent\hrule\hbox{\vrule\hskip 3pt
\vbox{\vskip 3pt
\hbox{#1}\vskip 3pt}\hskip 3pt\vrule}\hrule}}

\def\references{\sectnonumber References\par
     \begingroup \frenchspacing
      \def\par{\endgraf\hangindent24pt}
       \parindent0pt \vskip-10pt}
\def\endreferences{\endgroup}
\overfullrule 0pt

\title{Binary-Induced Gravitational Collapse:
       A Trivial Example}

\author{Stuart L. Shapiro}

\address{Department of Physics}
\address{Loomis Laboratory of Physics}
\address{University of Illinois at Urbana-Champaign}
\address{1110 West Green Street}
\address{Urbana, IL 61801-3080}
\address{and}
\address{Department of Astronomy and}
\address{National Center for Supercomputing Applications}
\address{University of Illinois at Urbana-Champaign}

\maketitle

\begin{abstract} We present a simple model illustrating how
a highly relativistic, compact object which is stable in
isolation can be driven dynamically unstable by the tidal
field of a binary companion. Our compact object consists of a
test-particle in a relativistic orbit about a black hole;
the binary companion is a distant point mass. Our example is
presented in  light of mounting theoretical opposition to
the possibility that sufficiently  massive, binary neutron
stars inspiraling from large distance can collapse to form
black holes prior to merger. Our strong-field model suggests
that first order post-Newtonian treatments of binaries,  and
stability analyses of binary equilibria  based on
orbit-averaged, mean gravitational fields, may not  be
adequate to rule out this possibility.
\end{abstract}

\vskip2pc]

\section{Introduction}

Binary neutron stars are among the most promising sources
for gravitational wave detectors now under construction,
like LIGO, VIRGO and GEO. This fact has motivated an intense
theoretical effort to understand their dynamical evolution
and predict the gravitational waveforms emitted during their
inspiral and coalescence.

Fully general relativistic treatments of the binary problem
are nontrivial because of the nonlinearity of Einstein's
field equations and the need for very large computational
resources to solve them, together with the equations of
relativistic hydrodynamics, in three spatial dimensions plus
time.  Recently, Wilson, Mathews and Maronetti (hereafter
WWM; ~\cite{WWM}) performed approximate relativistic
simulations which suggest that neutron stars just below the
maximum allowed rest mass and dynamically stable in
isolation become unstable to collapse to black holes  when
placed in a close binary orbit. Their results are in
disagreement with earlier  calculations of stability
performed in Newtonian  gravitation, which show that tidal
fields stabilize binary stars against radial collapse
~\cite{LAI}. The WWM results are also surprising in light of
several recent, albeit simplified, post-Newtonian (PN)
perturbation analyses ~\cite{LAI2}~\cite{LOM}, none of which
indicate the presence of any  relativistic radial
instability in binaries, at least to first  post-Newtonian
(1PN) order.

To help clarify the issue, Baumgarte et. al. ~\cite{TWB}
performed fully relativistic numerical calculations of binary
neutron stars in quasi-equilibrium circular orbits. (The
solutions correspond to ``quasi"-equilibrium because of the
inevitable slow, secular decay of the orbit due to the
emission of gravitational waves). They found that the
maximum allowed rest mass (i.e. baryon mass) of a neutron
star in a synchronous binary is in fact slightly larger
than  in isolation.  PN treatments of the same problem
confirm this conclusion  and extend it to nonsynchronous
binaries as well (e.g. Ref ~\cite{LOM}).  This finding rules
out at least one explanation for the results of WWM --
namely, the possibility that binary  equilibrium solutions
might not exist for neutron stars  with the highest values of
rest mass allowed for isolated stars. Using their
relativistic models,  Baumgarte et. al. ~\cite{TWB2} further
showed that all stars which are stable to radial collapse in
isolation are also  stable when placed in a binary, at least
to secular modes,  all the way down to the innermost stable
circular orbit.  This result, together with the expectation
that dynamical instabilities in binaries always arise at
smaller separation than secular instabilities, as in
Newtonian theory ~\cite{CC}~\cite{LAI}, suggests that a
dynamical instability against radial collapse is just not
present for binary neutron stars that inspiral from large
distance. This conclusion has been considerably strengthened
by an analytic, relativistic analysis by Thorne ~\cite{KST},
who argues  that a rigidly rotating, stable neutron star in
an inspiraling binary system will be protected against
secular and dynamical collapse by tidal interaction with its
companion.

It would appear that the WWM result may not be correct and
that their finding  might be the result of inadequate
computational resolution or one or  more of the
simplifications they adopt to solve this difficult numerical
problem. Nevertheless, their work has raised the very
interesting and nonobvious question about whether or not
tidal fields  in a relativistic binary can under ${\it any}$
circumstances trigger the collapse of a compact object that
is known to be stable in isolation.

In this paper, we offer a simple example to illustrate the
possibility of binary-induced dynamical collapse of a
compact object. The compact object consists of a test
particle in a tight orbit around a black hole. This
``compact object" is itself placed in orbit about a distant
mass and the ``binary'' -- i.e., the ``compact object'' in
circular orbit about the distant mass -- is allowed to
inspiral by the emission of gravitational radiation. We
investigate whether the tidal field of the distant mass can
drive such a compact object dynamically unstable even
though the object, when in isolation, is stable. Here,
dynamical instability manifests itself as a rapid plunge of
the test particle toward the black hole.

Our analysis of this model is highly simplified and arguably
heuristic in parts. Nevertheless, we are confident that our
overall picture is reliable and that our essential
conclusions will stand up to more rigorous scrutiny.

\section{Basic Model and Equations}

Consider at first a self-gravitating, N-body system in the
post-Newtonian limit of general relativity. The (geodesic)
equations of motion for body
$i$ may be cast in the form

\begin{equation} 
\label{one}
\bf {\ddot r}_i = -{\sum_{j \neq i} {m_j
\over r_{ij}^3} {\bf r_{ij}}} + {\bf F_i^{({\rm
PN})}}(1,2,...,i,...,N),\end{equation} 
where $m_j$ is the
total mass-energy of body $j$,
$\bf {r_j}$ is its coordinate position and $\bf
{r_{ij}=r_{i}-r_{j}}$. The first term on the right hand side
of Eq.~(\ref{one}) is the usual Newtonian (Coulomb) expression,
while the second term is the PN correction.  Explicit
expressions for this PN correction  have been developed by
many authors to various orders and in several gauges.
Einstein, Infield and Hoffman ~\cite{AE} provided the ``EIH"
equations of motion, which are valid to 1PN order. Damour
and Shafer ~\cite{TD} derived an N-particle Lagrangian valid
to 2PN order (see Ref ~\cite{TD2} for a review and
references). In these treatments, the PN correction to the
equation of motion for a particle typically depends on the
positions, velocities and the  accelerations of all of the
$N$ particles.

Let us now specialize to $N=3$.  Assume that bodies $1$ and
$2$ are appreciably closer to each other than they are to
body $3$,  and treat the influence of $3$ on the relative
orbit of the close pair $1-2$ as a small perturbation.
Define the quantities $m=m_1 + m_2$ and

\begin{equation}
\label{two}
{\bf R_{\rm{cm}}} \equiv {{m_1 {\bf r_1} +
m_2 {\bf r_2}} \over m},~~~ {\bf r \equiv r_1-r_2},~~~{\bf
r_{\rm{cm}} \equiv R_{\rm{cm}}-r_3},\end{equation} 
and use
Eq.~(\ref{one}) to write the equation of motion for the  relative
orbit of the close pair in the form

\begin{equation}
\label{three}
\ddot{\bf r}  =\ -\  {m \over r^3}\ {\bf
r}\ +\ {\bf f}_{\rm{(tidal)}}\ +\ {\bf F}_1^{\rm (PN)}
(1,2)\ -\ {\bf F}_2^{{\rm(PN)}}\ (1,2),
\end{equation}
where

\begin{equation}
\label{four}
{\bf f}_{\rm (tidal)}\ =\ {m_3 \over
r^3_{\rm cm}} \
\ \ \left[ {3{\bf r}_{\rm cm} \cdot {\bf r} \over r_{\rm cm}^2}
\ \ {\bf r}_{\rm cm}\ \ -\ \ {\bf r} \right]
\end{equation}
is the leading Newtonian tidal perturbation due to body $3$
and where the leading post-Newtonian corrections $\bf {F_1}$
and $\bf {F_2}$ depend  only on bodies $1$ and $2$. Here we
assume that $r \ll r_{cm}$ and we drop tidal corrections of
${\cal O}(m_3 {r^2\over r_{cm}^4})$. To the same order, the
orbit of particle $3$ about the  close pair is determined by

\begin{equation}
\label{five}
{\bf \ddot{r}}_{\rm cm}\ =\ -\ {M \over
r^3_{\rm cm}} {\bf r}_{\rm cm}, \end{equation} 
where
$M=m+m_3$ and where we drop tidal terms of
${\cal O}(M {m_< \over m_>} {r^2\over r_{cm}^4})$, as well
as the  very small PN corrections. Here $m_>\equiv
 max[m_1,m_2]$, and similarly for
$m_<$. To this order, the orbit of body $3$ about the close
pair decouples from the relative orbit of that pair.

In this perturbative treatment, let us henceforth examine
two extreme, opposite limits: In one limit, assume that the
orbital plane of body $3$ about the close pair moves in the
plane describing the relative orbit of that pair  (i.e., the
inclination angle of the two orbital planes satisfies $i =
0^o$). In the other limit, assume that $3$ moves in  a plane
perpendicular to the orbital plane of the pair  ($i =
90^o$).  Decompose Eqs.~(\ref{three}) and (\ref{five})  into radial and
angular components, yielding for the relative orbit of the
pair

\begin{equation}
\label{six}
\ddot{r} \ =\ -\ \left({m \over r^2}
\right)\ \ \left[ A\ +\ B\dot{r}\right]\ +\ r\dot{\phi}^2\ +\
f^{\hat{r}}_{{\rm (tidal)}},\end{equation}

\begin{equation}
\label{seven}
\ddot{\phi}\ =\ -\ \dot{\phi}
\left[\left({m\over r^2}\right) B\ +\ 2\ {\dot{r}\over r}
\right]\ +\  {1
\over r}\ f^{\hat{\phi}}_{{\rm (tidal)}}, \end{equation}
where

\begin{equation}
\label{eight}
f^{\hat{r}}_{{\rm (tidal)}}\ =\
\cases{\left({m_3 r
\over r^3_{{\rm cm}}}
\right)\left[3\ cos^2 (\phi - \theta) -1 \right],\
\ \ \ \ (i\ =\ 0^\circ )
\cr   \left({m_3 r\over r^3_{{\rm cm}}}\right)\left[3\ cos^2
(\theta)\ cos^2 (\phi) -1 \right] ,
\ \ \ \ (i \ =\ 90^\circ )} \end{equation}

\begin{equation}
\label{nine}
f^{\hat{\phi}}_{{\rm (tidal)}}\ =\
\cases{\left(-\ {m_3 r \over r^3_{{\rm cm}}}
\right)\left[3\ cos (\phi - \theta)\ sin(\phi - \theta)
\right],\
\ \ \ \ (i\ =\ 0^\circ )\cr
\left(-\ {m_3 r\over r^3_{{\rm cm}}}\right)\left[3\ cos^2
(\theta)\ cos(\phi)\ sin(\phi) \right],
\ \ \ \ (i \ =\ 90^\circ )} \end{equation} and for the orbit
of
$3$ about the pair

\begin{equation}
\label{ten}
\ddot{r}_{{\rm cm}}\ =\ -\ {M \over
r^2_{{\rm cm}}}\ +\ r_{{\rm cm}}\dot{\theta}^2,
\end{equation}

\begin{equation}
\label{eleven}
\ddot{\theta}\ =\ -\
{2\dot{\theta}\dot{r}_{{\rm cm}}
\over r_{{\rm cm}}}. \end{equation} In the Newtonian limit,
$A=1$ and $B=0$ in Eqs.~(\ref{six}) and (\ref{seven}).  In the case of
isolated binaries, Lincoln and Will ~\cite{CWL} derive
post-Newtonian expressions for $A$ and $B$ for arbitrary
masses, correct through 2.5PN order.  Kidder, Will and
Wiseman (hereafter KWW; ~\cite{LEK})  provide a ``hybrid" set
of equations in which the sum of the terms in $A$ and $B$
that are independent of the ratio
$\eta = \mu /m, \mu = m_1 m_2/m,$ is replaced by the exact
expression for geodesic motion in the Schwarzschild metric
of a body of mass $m$, while  the terms dependent on
$\eta$ are left unaffected. Their resulting equation  of
motion is therefore exact in the test-body limit $(\eta
\rightarrow 0)$  and is valid to 2.5PN order when
appropriately expanded for arbitrary masses. We shall
utilize these same hybrid expressions for $A$ and $B$ and,
for  simplicity, work in the test-body limit by taking one
member of our close pair to have a mass much smaller than
the other. Adopting harmonic (or de Donder) coordinates, the
resulting (Schwarzschild) expressions for $A$ and $B$ are
given by

\begin{equation}
\label{twelve}
A \ =\ \left[{1-m/r\over (1\ +\ m/r)^3}
\right]\ \ -\ \
\left[ {2-m/r \over 1-(m/r)^2}\right]\ {m \over r}\
\dot{r}^2\ +\ v^2 ,\end{equation}

\begin{equation}
\label{thirteen}
B\ =\ -\ \left[{4-2m/r \over
1-(m/r)^2}\right]\ \dot{r}\
\ \ \ ,\end{equation} 
where

\begin{equation}
\label{fourteen}
v^2\ =\ \dot{r}^2\ +\ r\dot{\phi}^2
.\end{equation}

We are interested in the dynamical behavior of the close
pair, regarded as a single ``compact object'', as it
inspirals toward the distant mass $m_3$ due to gravitational
radiation emission.  To treat the inspiral of this
``binary'' ($m_3$ in orbit about the ``compact object''),
we must include radiation reaction terms in the lowest order
(Newtonian) orbit Eqs.~(\ref{ten}) and (\ref{eleven}).  Formally, such a
treatment requires a consistent expansion up to 2.5PN
order.   In lieu of this, we shall analyze the inspiral by
assuming that the binary is in a nearly circular, Keplerian
orbit, which undergoes a slow inspiral due to gravitational
radiation loss in the quadrupole limit. This assumption is
equivalent to inserting a quadrupole radiation reaction
potential in the binary orbit Eq.~(\ref{five}) and neglecting the
lower-order, (non-dissipative)  PN corrections in that
particular equation. While such an expression is not
formally consistent to 2.5PN order, we believe that it
faithfully tracts the  secular inspiral of the binary in the
limit treated here in which the binary system is at wide
(nonrelativistic) separation. The details of  the inspiral
are not important here, only that the inspiral serves to
bring a tidal perturber slowly in from infinity toward our
compact object ~\cite{OUR}.  The resulting equation for the
binary inspiral is then

\begin{equation}
\label{fifteen}
r_{\rm cm}(t)/r_{\rm cm}(0) = (1 -t/T)^{1/4}
\end{equation} and

\begin{equation} 
\label{sixteen}
\theta(t) -\theta(0) ={4 \over 3}\left({ M
\over r_{\rm cm}(0)^3}
\right)^{1/2}T [1 - (1-t/T)^{3/4}], \end{equation} 
where the
binary inspiral timescale $T$ is given by

\begin{equation}
\label{seventeen}
T/m = {5 \over 256} {(r_{\rm cm}(0)/m)^4
\over {(M/m)(m_3/m)}}. \end{equation} 
We will use Eqs.
~(\ref{fifteen})--(\ref{seventeen}) in analyzing the relative orbit  Eqs.
~(\ref{six})--(\ref{nine}).

All orbital timescales in our problem are considerably
shorter than
$T$. For some purposes it is instructive to perform an orbit
average of the tidal terms appearing in Eq.~(\ref{six}) and (\ref{seven}), 
which yields
$\langle {f}^{\hat r}_{\rm tidal}\rangle = \alpha m_3
r/r_{cm}^3$, where $\alpha = 1/2$ for $i=0^o$, --1/4 for
$i=90^o$, and
$\langle {f}^{\hat \phi}_{\rm tidal}\rangle = 0$.

\section{Quasi-Equilibrium and Stability:  Mean-Field Approach}

Use the orbit-averaged tidal force to consider the close
pair  as a compact object in the  mean tidal field of its
distant, binary companion. In this approximation the mean
field, quasi-equilibrium state for the compact object object
consists of a test particle in  circular orbit at constant
radius about a black hole. The degree of compaction of this
quasi-equilibrium compact object is parametrized by the
radius of the circular orbit,
$r_{0}/m$; for small values of this parameter $\lesssim 10$,
the object is relativistic.  The quasi-equilibrium orbit in
the mean tidal field is  determined self-consistently by
requiring $\dot{r} = 0 = \ddot{r} =
\ddot{\phi}$. Eq.~(\ref{six}) then implies the circular orbit
condition

\begin{equation}
\label{eighteen}
\Omega^2_0\ \equiv\ \dot{\phi}^2_0\ =\ {mA_o
\over r_0^3}\ -\
\alpha\ {m_3 \over r^3_{{\rm cm}}}, \end{equation} 
which can
be solved along with Eqs.~(\ref{twelve}) and (\ref{fourteen}) 
to get $\Omega_0$
as a  function of $r_0$.  Because the orbiting particle is a
test-particle of negligible mass, its orbit does not decay
due to gravitational radiation, so the compact object is
secularly stable. Dynamical stability of the circular orbit
is determined by considering linear perturbations about the
equilibrium values  of $r_0$,
$\dot{\phi_0}$ and $\dot{r_0}=0$. The resulting stability
analysis of our compact object is a simple extension of the
analysis by  KWW for PN binaries in the absence of a tidal
field.  We find by linearizing Eq.~(\ref{six}) in a mean tidal
field that the condition for stable circular orbits may
again  be written as $a+bc < 0$, where now

\begin{equation}\matrix{& a &\ =\ & 3\Omega^2_0\ -\ \left({m
\label{nineteen}
\over r_0^2}\right)\ \
\left( {\partial A \over \partial r}\right)_0\ \ +\ \
3\alpha {m_3 \over r^3_{{\rm cm}}}\  \cr & b &\ =\ &
2r_0\Omega_0\ -\ \left({m \over r_0^2}\right)\ \
\left({\partial A \over \partial \Omega}\right)_0 \ \ \ \
\ \ \ \ \ \
\ \ \
\cr & c &\ =\ &-\Omega_0\ \left[ {2 \over r_0}\ +\
\left({m \over r^2_0}\right)\ \left({\partial B \over
\partial \dot{r}}\right)_0
\right]\ \ \ \ \ \ \ \ \ \ \ \ \  \cr} \end{equation} 
We can
analyze some limiting cases analytically. In the absence of
a tidal field ($r_{\rm cm}/m \rightarrow \infty$),  the
condition for dynamical stability reduces to

\begin{equation}
\label{twenty}
{r_0 \over m}\ >\ 5\ \ \ ,\ \ \Omega^2_0m^2\
\ <\ \ {1
\over 6^3}\
\ \ \ (r_{\rm cm}/m \rightarrow \infty). \end{equation} 
Eq.~(\ref{twenty}) reproduces the well-known value (in harmonic
coodinates) of the  innermost stable circular orbit (ISCO)
in Schwarzschild geometry,  $r_0/m=5$. Stability for a
strictly Newtonian orbit ($r_0/m \gg 1$) in the presence of
a tidal field from a perturber in the orbital plane requires

\begin{eqnarray}
\label{twenty-one}
r_0/r_{{\rm cm}}\ \ <\ \ \left({1 \over 2}\
\ {m\over m_3}\right)^{1/3}\ ,\ \ \Omega^2_0\ \ >\ \
{2m_3\over r^3_{{\rm cm}}} \\  \nonumber
\ ({\rm Newtonian},~~ i\ =\ 0^{\circ} ), \end{eqnarray}
which is the familiar Newtonian condition for avoiding tidal
disruption. For a tidal perturber perpendicular to the
orbital plane, all Newtonian orbits are stable.

\begin{figure}
\epsfxsize=3in
\begin{center}
\leavevmode
\epsffile{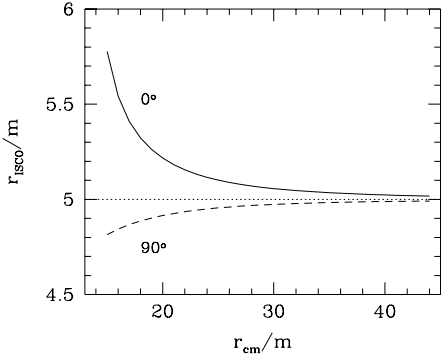}
\end{center}
\caption{The innermost stable circular orbit (ISCO)
as a function of the separation between the compact object
and its binary companion in the mean-field approximation.
The solid line is for coplanar test-particle and binary
companion orbits ($i=0^o$); the dashed line is for
orthogonal orbits ($i=90^o$).  The dotted line shows the
ISCO for an isolated compact object.}
\end{figure}

\begin{figure}
\epsfxsize=3in
\begin{center}
\leavevmode
\epsffile{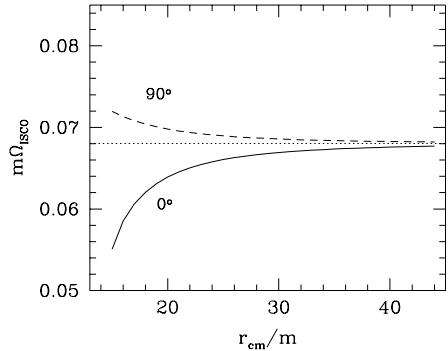}
\end{center}
\caption{ Angular frequency at the ISCO as a function
of the  separation between the compact object and its binary
companion in the mean-field approximation. Curves are
labeled as in Fig 1.}
\end{figure}

The more interesting cases involve the effect of the mean
tidal field of a companion on the stability of a
relativistic compact  object in binary equilibrium. To
analyze these,  we numerically evaluate  Eq.~(\ref{nineteen}) to
determine the ISCO as a function of the  strength of the
tidal field; results are summarized in Fig 1.   We set
$m=m_3$ so that the compact object and its distant companion
have equal mass. We see that  for a perturber in the orbital
plane of the close pair, the ISCO moves out as the perturber
approaches and the tidal field increases. Sufficiently tight
orbits  with $r_0/m>5$, which are always stable when the
compact object in isolation, are destabilized by the
companion. To guarantee that this result is not just a
coordinate effect, we can parametrize an orbit not only by
its radius, which is not gauge-invariant, but also  by its
period or angular frequency $\Omega_0$, which can be
measured by a distant observer. We see from Fig 2 that for a
perturber in the orbital plane, the angular frequency of the
ISCO moves to lower frequencies with increasing tidal  field
strength. Thus, a given compact object, specified by a fixed
angular frequency, can be stable in isolation but driven
dynamically unstable when in a binary system.

From Fig 1 we see, by contrast, that for a perturber
perpendicular to the orbital plane the ISCO actually moves
in. Likewise, we see from Fig 2 that the orbital frequency
at the ISCO increases. Hence, in this case the mean tidal
field due  to a companion would appear to stabilize the
compact object.

\section{Stability During Inspiral}

To assess the true fate of our compact object inspiraling
toward its binary companion, we abandon the mean-field
approximation and integrate the  relative orbit Eqs.
~(\ref{six})--(\ref{nine}) together with the inspiral Eqs. (\ref{fifteen})--
(\ref{seventeen}). At
$t=0$ we place a test-particle in a circular orbit about a
black hole, setting  $r/m = 5.9, {\dot r} = 0 =\phi$ and
${\dot \phi}$ given by Eq.~(\ref{eighteen}).  We set $m_3 = m$ and start
the companion $m_3$  at large separation $r_{\rm cm}/m =
40$. Hence, initially,  the tidal field of the companion is
negligible at the compact object and with so large a test
particle orbit,  the compact object begins in stable
equilibrium, almost exactly as it would be were it in
isolation.  We again treat two cases for the relative
orientation of the two orbital planes,
$i = 0^o$ and $i = 90^o$, setting $\theta = 0^o$ at $t=0$ in
the former case (companion aligned with the close pair) and
$\theta = 90^o$ in  the later case, (companion situated
above the orbital plane of the close pair, along its normal)
At such large binary separation, $\Omega_0 T
\gg 1$, hence the binary  orbits many times before the
inspiral proceeds very far. We thus expect  (and have
verified numerically) that   the fate of the compact object
does not depend on our choices of initial values of
$\phi$ and $\theta$.

\begin{figure}
\epsfxsize=3in
\begin{center}
\leavevmode
\epsffile{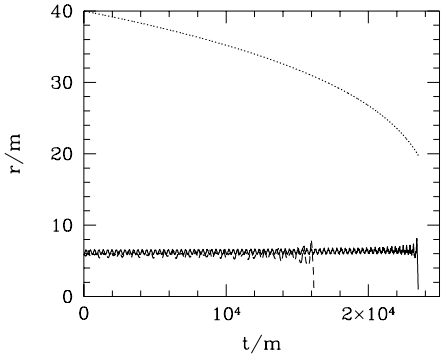}
\end{center}
\caption{ The orbital radius of the test particle
about the black hole as a function of time. The solid curve
shows the motion for coplanar test-particle and binary
companion orbits ($i=0^o$); the dashed curve shows the
motion for orthogonal orbits ($i=90^o$). The dotted curve
shows the separation between the compact object and its
binary companion as they inspiral together.}
\end{figure}

\begin{figure}
\epsfxsize=3in
\begin{center}
\leavevmode
\epsffile{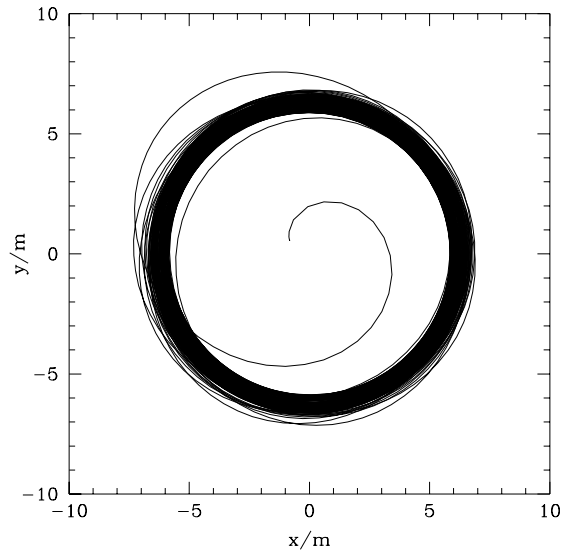}
\end{center}
\caption{ The trajectory of the test particle about
the black hole during binary inspiral for a companion moving
in a coplanar orbit ($i=0^o$).}
\end{figure}

\begin{figure}
\epsfxsize=3in
\begin{center}
\leavevmode
\epsffile{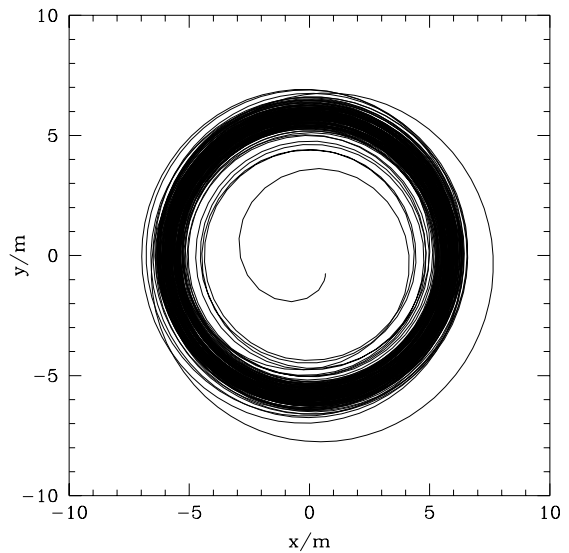}
\end{center}
\caption{ The trajectory of the test particle about
the black hole during binary inspiral for a companion moving
in an orthogonal plane ($i=90^o$).}
\end{figure}

The results of our integrations are summarized in Figs 3-5,
where we see that early on the compact object remains in
stable equilibrium for many  binary periods as the companion
slowly approaches. The test-particle orbit during this phase
is characterized by stable, small oscillations  about the
initial circular orbit; the oscillations occur at the beat
frequency between the distant binary companion and
test-particle orbits (cf. Eq~(\ref{eight})). When the companion gets
sufficiently close ($r_{\rm cm}/m \approx 20$ for $i=0^o$,
$r_{\rm cm}/m \approx 30$ for
$i=90^o$), the amplitude of the oscillations becomes
sufficiently large to cause the test-particle to plunge
toward  the hole and be captured. Hence in both cases the
tidal field of the companion  ultimately drives the
dynamical collapse of the compact object.

When $i=0^o$, this binary-induced collapse is  anticipated
from the mean-field stability analysis,  since  in this
case, the tidal field destabilizes close circular
test-particle orbits which are otherwise be stable in
isolation (see Sec 3). The unexpected destabilization in the
case  $i=90^o$ appears to result result from the growing
eccentricity of the test-particle orbit as the companion
approaches, eventually driving the particle too close to the
hole at pericenter and to ultimate capture.  In this case,
the mean-field analysis, based on orbit averaging and strict
circular motion, does not account for the more complicated
motion of the test particle and fails to predict its
ultimate fate.

\section{Discussion}

We have presented a simple illustration of a compact object
which is stable in isolation but dynamically unstable to
collapse when inserted in a binary. Admittedly, our compact
object is highly idealized and our treatment is simplified.
Our whole system consist of only 3 bodies moving on
geodesics. We could trivially generalize our compact object
to be a fully three-dimensional swarm of test particles, all
in randomly oriented circular orbits at the same fixed
radius about a central hole. Whenever  this radius is close
to, but slightly larger than 5M, then according to our
previous analysis, the presence of   a sufficiently close
binary companion would drive most, if not all, of the
test-particle orbits dynamically unstable during inspiral
from large separation. Once again, such a compact object
would be stable in isolation but could be driven dynamically
unstable when placed in a binary system.

It is not clear what bearing our simple example has to the
finding of WWM, especially in light of the counter evidence
suggesting that binary-induced collapse does not occur in
fluid stars. We have not incorported any hydrodynamic forces
in our model, and these forces are necessary to prevent
catastrophic collapse in fluid systems. It is interesting
nevertheless  that a simple collisionless particle model
exists which is stable in isolation but undergoes dynamical
collapse during binary inspiral. Typically, collisionless
equilibria in isolation  are subject to the many of the
same relativistic instabilities as fluid systems, including
radial instability to collapse,  (see, eg, Ref ~\cite{SLS}
for a review and references), so it is certainly relevant
for WWM and others  to have considered the possibility of
binary-induced collapse in the case of binary neutron stars.

Our simple model calculation offers a few cautions regarding
the treatment of the relativistic binary problem. Crucial to
our finding was our use of a high order relativistic
equation for the dynamical behavior of of our compact
object. Indeed, had we used 1PN expressions for $A$ and $B$
in Eqs.~(\ref{six}) and (\ref{seven}) instead of the hybrid expressions
Eqs. ~(\ref{twelve}) and (\ref{thirteen}) we would have  found spurious
behavior, because at 1PN (and even 3PN), the solution for
the test-particle ISCO in a PN expansion of the Schwarzshild
equations of motion yields a spurious root that does not
converge to 5m for isolated holes ~\cite{LEK}. By using the
hybrid expressions in Eqs.~(\ref{six}) and (\ref{seven}), our compact
object is treated exactly to all orders, at least when it is
isolation.

It is not surprising that a 1PN treatment is inadequate for
this problem. Such an approximation is not even sufficient
to determine reliably the onset of radial  instability to
collapse in an isolated fluid star. There the highly
nonlinear terms of Einstein's equations play a crucial role
and only for special equations of state (e.g
$n=3$ polytropes) is a 1PN analysis quantitatively reliable
~\cite{LOM}.

We have also seen that stability analyses of equilibria
based on orbit averaging and mean tidal fields  may prove
misleading, as they did for the $i=90^o$ case treated above.
By the same token, it is by no means clear that conclusions
found for synchronous stellar binaries will apply to
nonsynchronous binaries. (Here we recall that 
the viscosity in neutron star binaries is  not sufficiently large
to drive the system to 
corotation ~\cite{CSK}). As our calculations demonstrate,
once we relax dynamical constraints  and allow motion with
greater degrees of freedom, the final outcome of binary
inspiral can be quite different, even qualitatively.

There is mounting evidence which argues against the
possibility that sufficiently massive, highly compact
neutron stars in coalescing binaries can collapse to black
holes prior to merger.  However, we will be greatly
reassured when this conclusion is  finally corroborated by
detailed hydrodynamical calculations in  full general
relativity.

\bigskip

\acknowledgements 

It is a pleasure to thank T. Baumgarte for several
useful discussions.  This work has been supported in part
by  NSF Grant AST 96-18524 and NASA Grant NAGW-5-3420.


\begin{references}
\bibitem{OUR} Our inspiral approximation is performed in the
same  spirit as the hybrid approximation, where complete PN
consistency is sacrificed in order to insure  that the
equations recover some essential dynamical behavior  arising
at higher order.
\end{references}
\end{document}